\documentstyle[pra,aps,multicol,epsf]{revtex}

\begin{document}

\title{Critical Temperature Shift in Weakly Interacting Bose Gas}

\author{V.~A.~Kashurnikov$^{1}$, N.~V.~Prokof'ev$^{2}$, and 
B.~V.~Svistunov $^{2,3}$}

\address{
$^1$Moscow State Engineering Physics Institute,
 115409 Moscow, Russia \\
$^2$ Department of Physics, University of Massachusetts, 
Amherst, MA 01003, USA \\
$^3$ Russian Research Center ``Kurchatov Institute",
 123182 Moscow, 
Russia  \\
 }

\maketitle
\begin{abstract} 
With a high-performance Monte Carlo algorithm we study the
interaction-induced shift of the critical point in weakly
interacting three-dimensional $| \psi |^4$-theory 
(which includes quantum Bose gas). 
In terms of critical density, $n_c$, mass, $m$, interaction, $U$,
and temperature, $T$, this shift is universal: 
$\Delta n_c(T) = - C m^3 T^2 U$,
the constant $C$ found to be equal to 
$0.0140 \pm 0.0005$. For quantum 
Bose gas with the scattering length $a$ this implies 
$\Delta T_c/T_c = C_0 a n^{1/3}$, with $C_0=1.29 \pm 0.05$.
\end{abstract}


\begin{multicols}{2}
\narrowtext

The problem of interaction-induced shift of the critical temperature in a 
weakly interacting Bose gas has been attracting a lot of theoretical effort 
during recent years \cite{Stoof,Ceperley,Holzmann,Baym,Krauth}.
Apart from an apparent fundamental interest, the study of this
 problem has been strongly
stimulated by the experimental achievement of Bose-Einstein
 condensation in ultra-cold
gases \cite{BEC} and, correspondingly, the perspective of
 experimental verification of
theoretical predictions. The first experimental study of
 the critical temperature
of interacting Bose gas was reported recently for 
the $^4 \rm He$-Vycor system \cite{Reppy}.

Despite a considerable number of papers dealing with the issue, 
the answer for the critical temperature shift still 
remains a matter of controversy. As it follows from
simple analysis of dimensions, the result should have the form
\begin{equation}
{\Delta T_c \over T_c} = C_0 a n^{1/3} \; ,
\label{DT}
\end{equation}
where $C_0$ is a dimensionless constant.
There are reasonable physical arguments for $C_0$ to be
 positive \cite{Ceperley},
and most studies predict that $C_0>0$. However, there is a profound
discrepancy between different approaches in the value of $C_0$.
The range of variation of
different predictions is an order-of-magnitude large: $0.34$
 \cite{Ceperley},
$0.7$ \cite{Holzmann}, $2.3$ \cite{Krauth}, $2.9$ \cite{Baym},
 $4.66$ \cite{Stoof}.
As far as the experiment of Ref.~\cite{Reppy} is concerned, 
where the law (\ref{DT}) was clearly observed, it should be
 realized that the only
quantity that was really measured was the product $C_0 a$.
 Therefore, until the 
accurate value of the scattering length $a$ is available,
 it is impossible to 
make a reliable conclusion about $C_0$ \cite{comment}. 

Though Eq.~(\ref{DT}) formally looks like a perturbative correction 
in terms of the gas parameter $a n^{1/3}$, physically it is clear that this 
is not the case (cf., however, \cite{Krauth}), because in the thermodynamic 
limit any finite interaction, no matter how small, changes the universality
class of the phase transition: while the ideal Bose gas belongs to the 
universality class of the Gaussian complex-field model, the interacting 
system pertains to the $XY$-model universality class. Due to this
fact the first-principle analytic description of the interacting Bose gas 
at the critical point is rather difficult. To the best of our knowledge, 
up to now there was done only one {\it ab initio} Monte Carlo 
simulation of the interacting gas in the context of $\Delta T_c$ 
(Gr\"{u}ter, Ceperley, and Lalo\"{e}, Ref.~\cite{Ceperley}). 
An alternative Monte Carlo approach by Holzmann and Krauth \cite{Krauth} 
was based on a {\it hypothesis} that Eq.~(\ref{DT}) can be obtained in a 
(rather sophisticated) perturbative way, by simulating an ideal gas.

The goal of this Letter is to develop a numeric scheme for describing weakly
interacting Bose gas in the fluctuation region, which, in particular, could 
produce an accurate result for $C_0$.
The crucial point that renders this aim attainable is the universality 
of the long-wave behavior of weakly interacting 
$| \psi |^4$-theories in the fluctuation region at transition point: 
All these theories, no matter quantum or classical, 
continuous or discrete, lead to a generic long-wave Hamiltonian 
(cf., e.g., \cite{Baym}) \begin{equation}
H =  \int  \left\{ {1 \over 2m} |\nabla \psi|^2   
 +{U \over 2}  | \psi |^4 \right\} d {\bf r} \; .
\label{H}
\end{equation}
From this universality it follows that in {\it any} such system the shift of
the critical density (which is not sensitive to the ultra-violet cutoff of 
the theory) is given by the formula (from now on $\hbar = 1$)
\begin{equation}
\Delta n_c(T) = - C m^3 T^2 U \; ,
\label{dn}
\end{equation}
where $C$ is a universal constant. Then, the critical
 temperature shift (which is not
universal, being sensitive to the short-wave physics)
 can be obtained for each 
particular system from the obvious relation 
\begin{equation}
{\Delta T_c \over \Delta n_c} = - { d T_c^{(0)}(n) \over d n }  \;  ,
\label{dn_dT}
\end{equation}
where $T_c^{(0)}$ is the critical temperature of the corresponding
 ideal system.
From Eq.~(\ref{dn_dT}) one readily obtains for the quantum Bose gas
\begin{equation}
C_0 \approx 91.8 C  \;  .
\label{C_C0}
\end{equation}

The universality of the density shift (\ref{dn}) suggests that it can 
actually be calculated not in a continuous quantum system, where
 simulations 
are computationally expensive, but in a discrete classical model
\begin{equation}
H = -\sum_{<ij>} [\psi_i^* \psi_j + {\rm c.c.}] +
{U \over 2} \sum_i | \psi_i |^4 \;  ,
\label{H1}
\end{equation}
where $i$ and $j$ stand for the sites of 3D cubic lattice,
 $<ij>$ denotes 
nearest-neighbor sites, and $\psi_i$ is a complex variable. 
The long-wave behavior of the discrete system (\ref{H1}) is
 described by the Hamiltonian
(\ref{H}) with $m=1/2$. The advantage of utilizing model
 (\ref{H1}) for numeric
analysis is associated with the existence of a very powerful
 Worm algorithm for
simulating discrete classical systems of this type \cite{PS}. 

{\large \it Analysis of dimensions.} We start with rendering general
considerations leading to Eq.~(\ref{dn}). Consider the grand canonical 
counterpart of the Hamiltonian (\ref{H})
\begin{equation}
H' = \int \left\{ -{1 \over 2 m} |\nabla \psi|^2 + 
{U \over 2} | \psi |^4 - \tilde{\mu} | \psi |^2
\right\} d {\bf r} \; .
\label{psi_4}
\end{equation}
Here $\tilde{\mu}$ is an effective chemical potential [which,
 generally speaking,
does not coincide with the bare chemical potential of the
 original system for
which $H'$ plays the role of the effective Hamiltonian].
As the field $\psi$ describes only the long-wave modes of
 the original system, it is supposed 
to have some ultra-violet cutoff:
\begin{equation}
\psi({\bf r}) = \sum_{k<k_0} a_{\bf k} e^{i {\bf kr}} \; .
\label{cutoff}
\end{equation}
The choice of $k_0$ is arbitrary as long as $k_0$ is larger than
the momentum $k_c$ which characterizes the momentum region where 
harmonics are strongly coupled to each other ($k \gg k_c$ harmonics 
are almost free). If $k_0 \sim k_c$, then all three terms 
in Eq.~(\ref{psi_4}) are of the same
order, and we have the following relations
\begin{equation}
k_c^2/m \, \sim \, | \tilde{\mu} | \, \sim \, \tilde{n} U \; , 
\label{est1}
\end{equation}
\begin{equation}
\tilde{n} \sim \langle | \psi |^2 \rangle \sim \sum_{k<k_c} n_k \; , 
\label{tilde_n}
\end{equation}
where $n_k = \langle |a_{\bf k}|^2 \rangle $ and $\langle
 \ldots \rangle$ means 
statistical averaging.

According to Eq.~(\ref{tilde_n}), $\tilde{n} \sim k_c^3 n_{k_c}$.  
Since $k_c$ is a momentum separating strongly coupled
long-wave harmonics from ideal short-wave
ones, the order-of-magnitude estimate for $n_{k_c}$ may be
obtained from the ideal gas formula :
\begin{equation}
n_{k_c} \sim \frac{T}{k_c^2/2m-\tilde{\mu}}
 \sim \frac{T}{\tilde{\mu}} \; .
\label{rel2}
\end{equation}
Substituting this back to 
Eqs.~(\ref{est1})-(\ref{tilde_n}) we see that
\begin{equation}
k_c \sim m^2 T U \; , \; \; \; 
\tilde{n} \sim m^3 T^2 U \; .
\label{est2}
\end{equation}

Consider now the shift of the critical density 
\begin{equation}
\Delta n_c(T) = \sum_{\bf k} [ n_k^{\rm (crit)} - 
n_k^{\rm (crit,0)} ] \; ,
\label{shift}
\end{equation}
where $n_k^{\rm (crit)}$ and $n_k^{\rm (crit,0)}$ 
are the occupation numbers at the critical density
 (corresponding to a 
given temperature $T$)
for interacting and non-interacting system, respectively. 
The main contribution to the sum in 
Eq.~(\ref{shift}) comes from $k \sim k_c$, since at $k \gg k_c$
  the terms 
$n_k^{\rm (crit)}$ and $n_k^{\rm (crit,0)}$ are almost equal 
and compensate each other. At $k \sim k_c$ one can use the
 estimate (\ref{rel2}),
which means that $\Delta n_c \sim \tilde{n}$, and we arrive
 at Eq.~(\ref{dn}).

{\large \it Numeric procedure and results.} Our Monte Carlo
 scheme is based on the
Worm-algorithm simulation of the high-temperature expansions for
the two-point correlator $\langle \psi_i^* \psi_j \rangle$
  in the grand canonical
ensemble. (The description of this approach and performance tests 
see in Ref.~ \onlinecite{PS}).

To proceed, we need a formal definition of the critical point for 
finite system with linear dimension $L$, which, on one hand, is 
consistent with the thermodynamic limit $L \to \infty$, and, 
on the other hand, is convenient from the computational viewpoint. 
We adopt the following definition. 
By critical we understand the point where
the condensate density equals to
\begin{equation}
n_0^{\rm (crit)}(L) = T/(3.75 L) \; .
\label{crit}
\end{equation}
This definition is optimized for the ideal gas (minimal 
finite-size corrections for the total density). Nevertheless, 
it proved to be quite satisfactory for the interacting 
system as well, including the limit of strong interaction.

The classical model (\ref{H1}) possesses an obvious similarity property: 
The transformation $\psi \to \sqrt{T} \psi$, $U \to U/T$ reduces the 
problem to the corresponding $T=1$ case.
Hence, without loss of generality, we set $T=1$.

\vspace*{-0.4cm}
\begin{figure}
\begin{center}
\epsfxsize=0.42\textwidth
\hspace*{0.5cm} \epsfbox{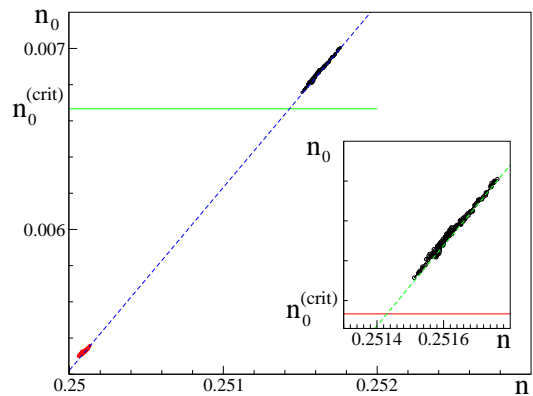}
\end{center}
\vspace*{-4.0cm}
\caption{ Co-variance of $n_0$ and $n$ for $U=1$, $L=40$,
 and two different values
of $\mu$. 
Data points correspond 
to successive Monte Carlo outputs $(n,n_0)$. The crucial
observation here is that fluctuations of $n_0$ and $n$ are
 strongly correlated,
and follow the line: 
$\delta n_0 / \delta n \approx d \bar{n}_0 / d \bar{n}$, where
bars above the symbols denote statistic limits. With changing
 chemical potential the
whole linear-shape pattern just shifts along its axis.
 We thus obtain
$n_c(U=1,L=40)=0.25143(3)$ from this plot.
}
\label{fig:fig1}
\end{figure}

The procedure of determining the critical density is as follows.
We iteratively tune the chemical potential $\mu$ until (\ref{crit}) 
is satisfied with a required accuracy, and then calculate the 
corresponding density. For small $U$, where a high 
precision is required to
resolve a small effect \cite{11}, we radically improve the efficiency of 
the pinpointing procedure
by utilizing the covariance (see Ref.~\onlinecite{Sandvick}) 
in the statistical fluctuations 
of Monte Carlo data for $n_0$ and $n$. 
The existence of strong covariance between fluctuations of 
$n_0$ and $n$ (Fig.~1) allows one to extract the dependence  
$n_0(n)$  with significantly higher
accuracy than the accuracy of separate calculations of
 $n_0(\mu)$ and $n(\mu)$.

\vspace*{-0.4cm}
\begin{figure}
\begin{center}
\epsfxsize=0.44\textwidth
\hspace*{0.5cm} \epsfbox{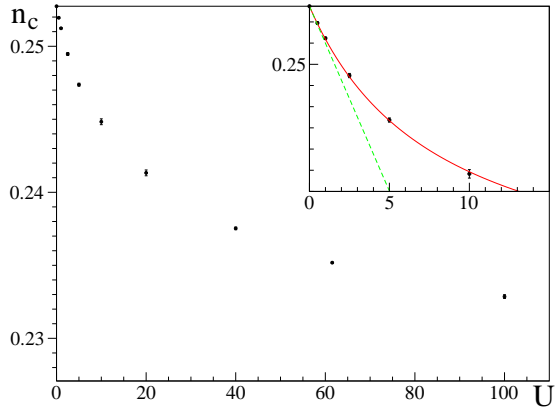}
\end{center}
\vspace*{-4.0cm}
\caption{Critical ratio $(n/T)_c$ for the model (\ref{H1}), 
as a function of interaction $U$. The upper bound of the box 
corresponds to $(n/T)_c = 0.25274$ of the ideal system, 
and the lower bound corresponds to the $U \to \infty$ limit, 
which is the $XY$-model with $(n/T)_c \approx 0.22708$. Insert:
The critical density at small $U$. 
The dashed line corresponds to Eq.~(\ref{dn}) with $C=0.0140$.
The solid line corresponds to more accurate  approximation
Eq.~(\ref{fit}) at $L \to \infty$:
$\Delta n_c=-Cm^3T^2U/(1 + b_0 U)$ with $b_0=0.123$.}
\label{fig:fig2}
\end{figure}

Prior to inspecting the region of small $U$, it is instructive to 
consider the behavior of the
critical density for the model (\ref{H1}) at finite interactions, in view of
a remarkable contrast to the quantum case. In a quantum gas with 
the parameter $an^{1/3}$ of order unity, the shift of the critical 
temperature is determined by a competition of two effects\cite{Ceperley}: 
(i) the suppression of the long-wave fluctuations of density 
(which stimulates condensation), and (ii) the depletion of the 
long-wave components of the field (which effectively reduces density, 
and thus drives the system away from the phase transition). 
The first circumstance dominates at small interaction, while for larger 
$an^{1/3}$ the second effect takes over. Following this competition, 
the shift of the critical temperature goes through a maximum 
around $an^{1/3} \lesssim 1$, and then goes down
finally becoming negative \cite{Ceperley,Reppy}. 
While the first effect is of classical-field nature and thus is
relevant to (\ref{H1}), 
the second effect is of quantum nature, and does not have any 
classical analog. That is why in the classical system we expect 
that the (negative) shift of the critical density will 
be a monotonous function of $U$. This is precisely what is 
revealed by our simulation (Fig.~2).
The critical density is monotonously decreasing with interaction, saturating
at $U \to \infty$ to the known value $(n/T)_c \approx 0.22708$
 corresponding to the 
$XY$-model (see, e.g., Ref.~\onlinecite{XY}).  
[As an illustration of the fact that the criterion (\ref{crit}) works 
reasonably well for the strongly interacting system, 
we reproduced the above-mentioned critical point of the $XY$-model with 
the accuracy of $5$ significant digits by the brut-force simulation 
of large systems with $L=160$, and checked that finite-size systematic 
errors are almost irrelevant already for $L=40$.]

Now we turn to the region of small $U$. As the accuracy of 
calculations can be significantly
improved by a proper finite-size analysis of the data, 
we begin with generalizing  relation (\ref{dn}) to the finite-size case. 
We note that in the limit $U \to 0$ the only relevant
lengthscale is the mode-coupling radius
\begin{equation}
r_c = {2\pi \over k_c} \sim {1 \over m^2 T U} \; ,
\label{rc}
\end{equation}
Hence, $r_c$ defines the characteristic size of the system that can be 
considered as macroscopic with respect to (small) interaction: 
The system is in the proper thermodynamic limit if, and only if $L \gg r_c$.
In the limit $L \ll r_c$ interactions can be treated perturbatively,
and thus are essentially irrelevant. From this consideration we 
arrive at the following generalization of Eq.~(\ref{dn}):
\begin{equation}
\frac{\Delta n_c(T,L)}{m^3T^2 U} = 
- f(r_c/L) + {\cal O} (U) \; , 
\label{main}
\end{equation}
where the particular form of the dimensionless function $f$ depends on the
definition of $n_c(L)$, apart from universal limits 
$f(0)=C$ and $f(\infty)=0$. On the basis of Eq.~(\ref{main}), 
we employ the following scaling function which should
adequately describe the limit $U \to 0$, $L \to \infty$ 
\begin{equation}
- \frac{\Delta n_c(T,L)}{m^3T^2 U} = 
\frac{C}{1+b_0U+(a_1+b_1 U)x^{\eta} } \; , 
\label{fit}
\end{equation}
where $x=1/(L m^2 T U)$ and $\eta=1.038$ \cite{XY} in accordance with
expected asymptotic behavior of the finite-size corrections for 
$XY$-type models. Fitting parameters $C$, $b_0$, $a_1$, and $b_1$ 
were obtained by stochastic optimization ($C=0.0140$, $a_1=1.29$,  
$b_0=0.123$, $b_1=0.744$). As is seen from Fig.~3, the scaling
function (\ref{fit}) works extremely well. With errorbars defined from
the uncertainty of the fitting procedure, for the constant $C$ we have 
\begin{equation}
C=0.0140 \pm 0.0005 \; ,
\label{C}
\end{equation}
which, according to Eq.~(\ref{C_C0}), means
\begin{equation}
C_0=1.29 \pm 0.05 \; . 
\label{Bose}
\end{equation}

\vspace*{-1.7cm}
\begin{figure}
\begin{center}
\epsfxsize=0.515\textwidth
\hspace*{0.cm} \epsfbox{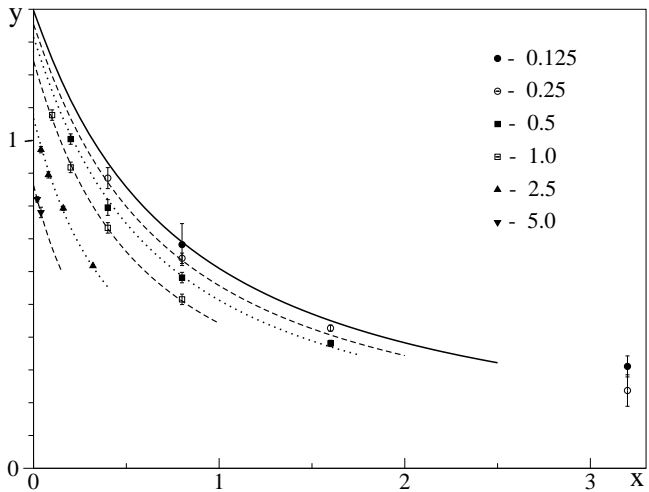}
\end{center}
\vspace*{-4.4cm}
\caption{Scaling plot of $y=10^2~\Delta n_c / (m^3T^2 U)$ vs $x=1/(L m^2 T U)$
at small $U$.
The upper curve is given by the scaling function at $U=0$. Note that while this
function is not supposed to fit the data at $x \ge 1$, actually it
is rather accurate up to $x \sim 3$.}
\label{fig3}
\end{figure}

While the particular form of $f(x)$ in the region $0 < x < \infty$ 
is sensitive to the definition of $n_c(L)$, the typical scale  $x \sim x_c$ 
of significant variation of $f(x)$ from its limiting value $f(0)$ 
depends only on the mode-coupling radius:
$x_c \sim 1/(r_c m^2 T U)$. From Fig.~3 we deduce
\begin{equation}
r_c \approx  1/(m^2 T U) \; . 
\label{rc_acc}
\end{equation}
Using Eq.~(\ref{rc_acc}), one can readily estimate how small should 
be the interaction for a particular weakly interacting system to demonstrate
the universal long-wave behavior. The generic requirement is that 
$r_c$ be much larger than other microscopic lengthscales at which 
the behavior of the system becomes
non-universal (due to the discreteness of the system, or/and
quantization of occupation numbers). The most severe requirement for
the quantum Bose gas is $r_c \gg \lambda_T$, where 
$\lambda_T \approx n^{-1/3}$ is the thermal de Broglie wavelength at the Bose
condensation point. In terms of the gas parameter this means 
$an^{1/3} \ll 0.025 $, and we immediately realize that neither Monte Carlo
simulation of Ref.~\onlinecite{Ceperley}, nor the experiment of
Ref.~\onlinecite{Reppy} have reached the universality region:
In both cases the minimal value of $na^3$ was $\sim 10^{-5}$,
while our estimate shows that one may expect universality only
at $na^3\lesssim 10^{-7}$.

In conclusion, we have developed a Monte Carlo approach 
for the study of weakly interacting $| \psi |^4$-theories 
in the fluctuation region. With this technique we obtained
an accurate result for the critical temperature of weakly interacting
three-dimensional Bose gas, and established a criterion of 
its applicability. We note that the method can be also applied
to weakly interacting Bose gas in two dimensions. The CPU time 
required for collecting high-precision data reported above
is equivalent to two years on PIII-733 processor. 

We are grateful to J.~Machta for his interest and discussions.
This work was supported by the National Science Foundation under Grant
DMR-0071767.

\end{multicols}

\begin{thebibliography}{99}
\bibitem{comment} The authors of Ref.~\cite{Reppy} estimate 
$C_0$ by adopting $a=0.22~{\rm nm}$, following Ref.~\cite{Runge}, 
which is appropriate for describing properties of solid $^4 \rm He$. 
The reliability of this assumption for the  $^4 \rm He$-Vycor system
is questionable, given completely different kinematics at 
the atomic lengthscale.
\end{thebibliography}
\end{document}